\documentclass[aps,prb,twocolumn,showpacs,eqsecnum]{revtex4}
\usepackage{bm}

\usepackage{amsmath}
\usepackage{amssymb}
\usepackage{graphicx}

\newcommand{\beq}{\begin{equation}}
\newcommand{\eeq}{\end{equation}}
\newcommand{\beqar}{\begin{eqnarray*}}
\newcommand{\eeqar}{\end{eqnarray*}}

\newcommand{\ua}{\uparrow}
\newcommand{\da}{\downarrow}

\newcommand{\dt}{{\text d}}
\newcommand{\etxt}{{\text e}}

\newcommand{\itxt}{{\text i}}

\newcommand{\Hh}{\hat{H}}

\newcommand{\psih}{\hat{\psi}}

\newcommand{\sigb}{{\mbox{\boldmath{$\sigma$}}}}

\newcommand{\rtarr}{\rightarrow}

\newcommand{\et}{\tilde{\epsilon}}

\newcommand{\Det}{\tilde{\Delta}}

\newcommand{\Eb}{{\bf E}}
\newcommand{\Hb}{{\bf H}}
\newcommand{\nb}{{\bf n}}

\newcommand{\kb}{{\bf k}}

\newcommand{\jb}{{\bf j}}

\newcommand{\s}{{\bf s}}

\newcommand{\pb}{{\bf p}}

\newcommand{\Ab}{{\bf A}}

\newcommand{\rb}{{\bf r}}

\newcommand{\dg}{\dagger}

\newcommand{\lan}{\langle}
\newcommand{\ran}{\rangle}
\newcommand{\om}{\omega}

\newcommand{\al}{\alpha}

\newcommand{\ga}{\gamma}

\newcommand{\de}{\delta}
\newcommand{\De}{\Delta}
\newcommand{\la}{\lambda}

\newcommand{\sig}{\sigma}

\newcommand{\vphi}{\varphi}

\newcommand{\e}{\epsilon}

\newcommand{\lt}{\left}
\newcommand{\rt}{\right}
\newcommand{\Qbr}{\bar{Q}}

\newcommand{\bwt}{\begin{widetext}}
\newcommand{\ewt}{\end{widetext}}

\begin{document}

\title{Surface impedance of superconductors with magnetic impurities}
\author{Maxim Kharitonov$^1$, Thomas Proslier$^2$, Andreas Glatz$^2$, and Michael J. Pellin$^2$}
\address{
$^1$Center for Materials Theory,
Rutgers University, Piscataway, NJ 08854, USA\\
$^2$Materials Science Division, Argonne National Laboratory, Argonne, IL 60439, USA}
\date{\today}
\begin{abstract}

Motivated by the problem of
the residual surface resistance
of the superconducting radio-frequency (SRF) cavities,
we develop a microscopic theory of the surface impedance of $s$-wave superconductors with magnetic impurities.
We analytically calculate the current response function and surface impedance
for a sample with spatially uniform distribution of impurities,
treating magnetic impurities in the framework of the Shiba theory.
The obtained  general expressions hold in a wide range of parameter values,
such as temperature, frequency, mean free path, and exchange coupling strength.
This generality, on the one hand,
allows for direct numerical implementation of our results
to describe experimental systems (SRF cavities, superconducting qubits)
under various practically relevant conditions.
On the other hand, explicit analytical expressions can be obtained in a number of limiting cases,
which makes possible further theoretical investigation of certain regimes.
As a feature of key relevance to SRF cavities,
we show that in the regime of ``gapless superconductivity''
the surface resistance exhibits saturation at zero temperature.
Our theory thus explicitly demonstrates that magnetic impurities,
presumably contained in the oxide surface layer of the SRF cavities,
provide a microscopic mechanism for the residual resistance.

\end{abstract}
\maketitle

\section{Introduction}

\begin{figure*}
\centerline{
\includegraphics[width=.70\columnwidth]{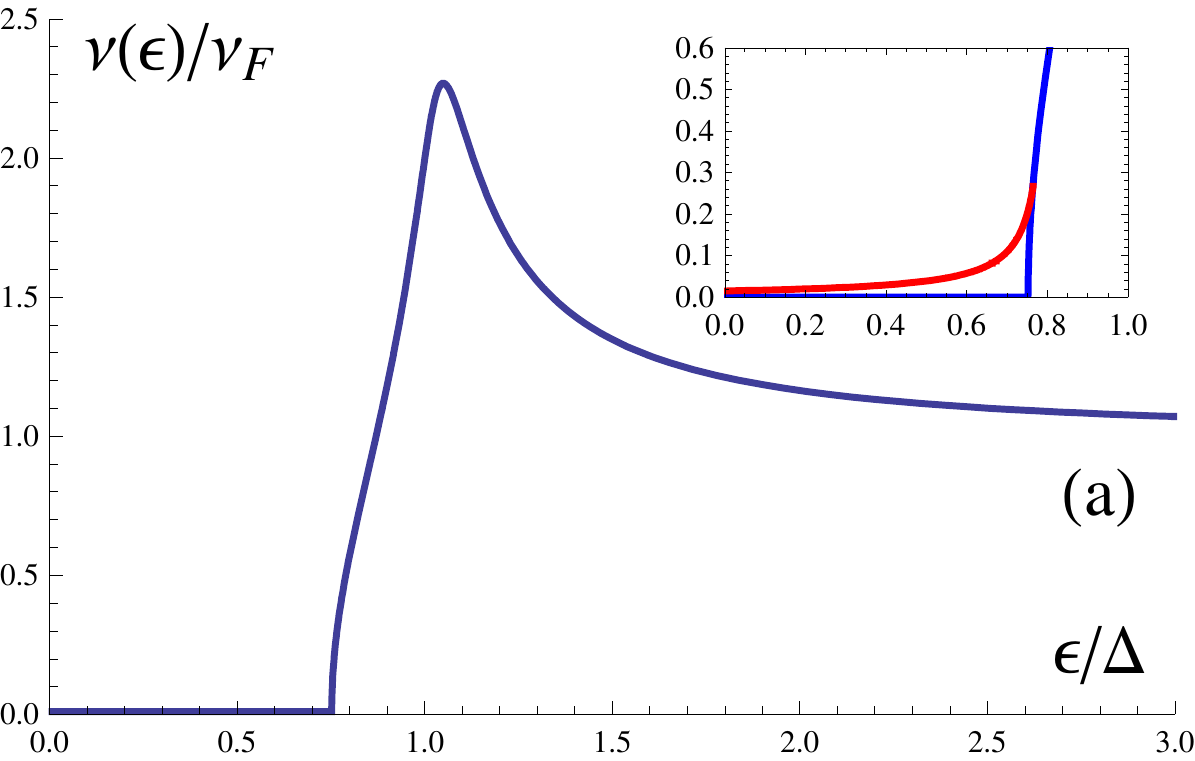}
\hspace{9mm}
\includegraphics[width=.72\columnwidth]{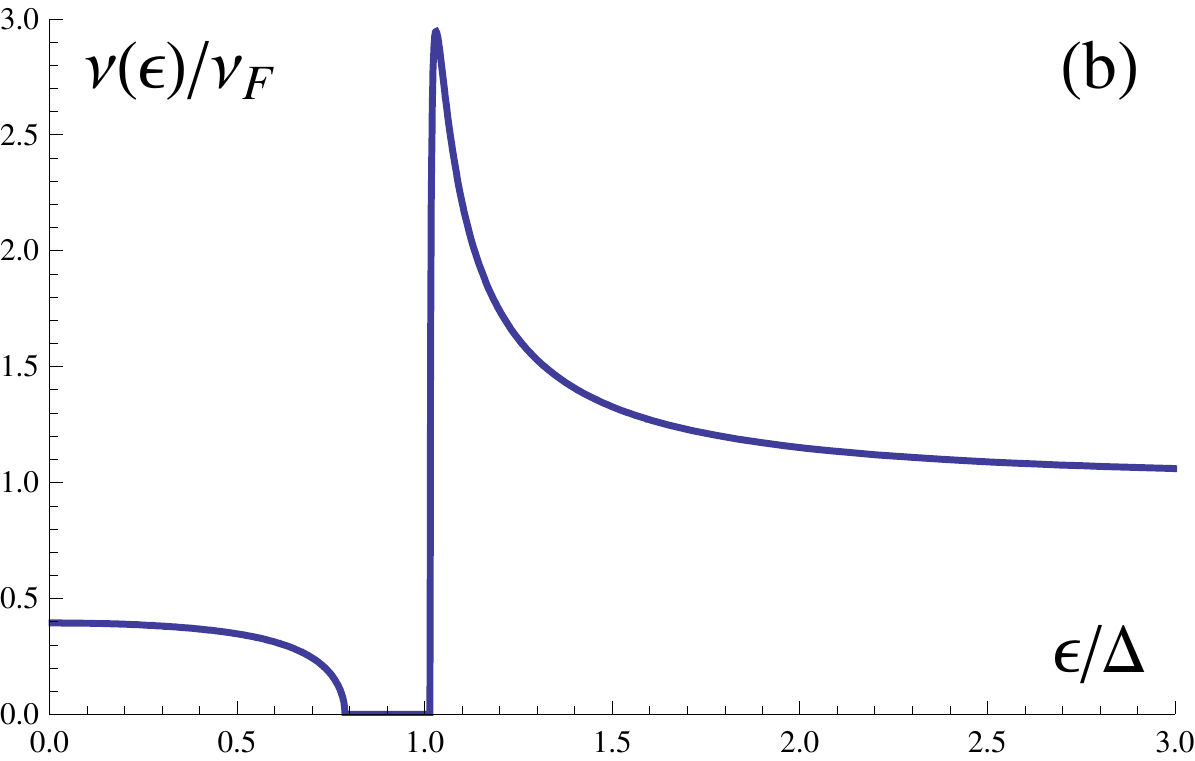}
}
\centerline{
\includegraphics[width=.70\columnwidth]{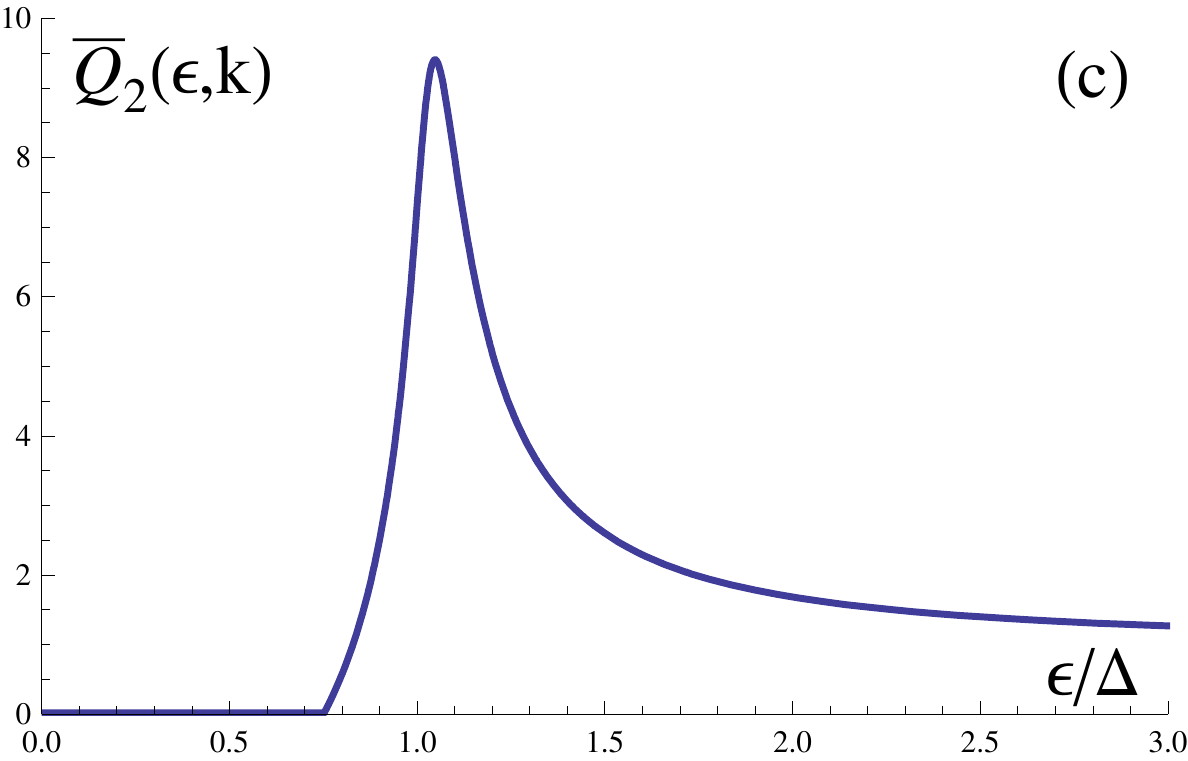}
\hspace{1cm}
\includegraphics[width=.70\columnwidth]{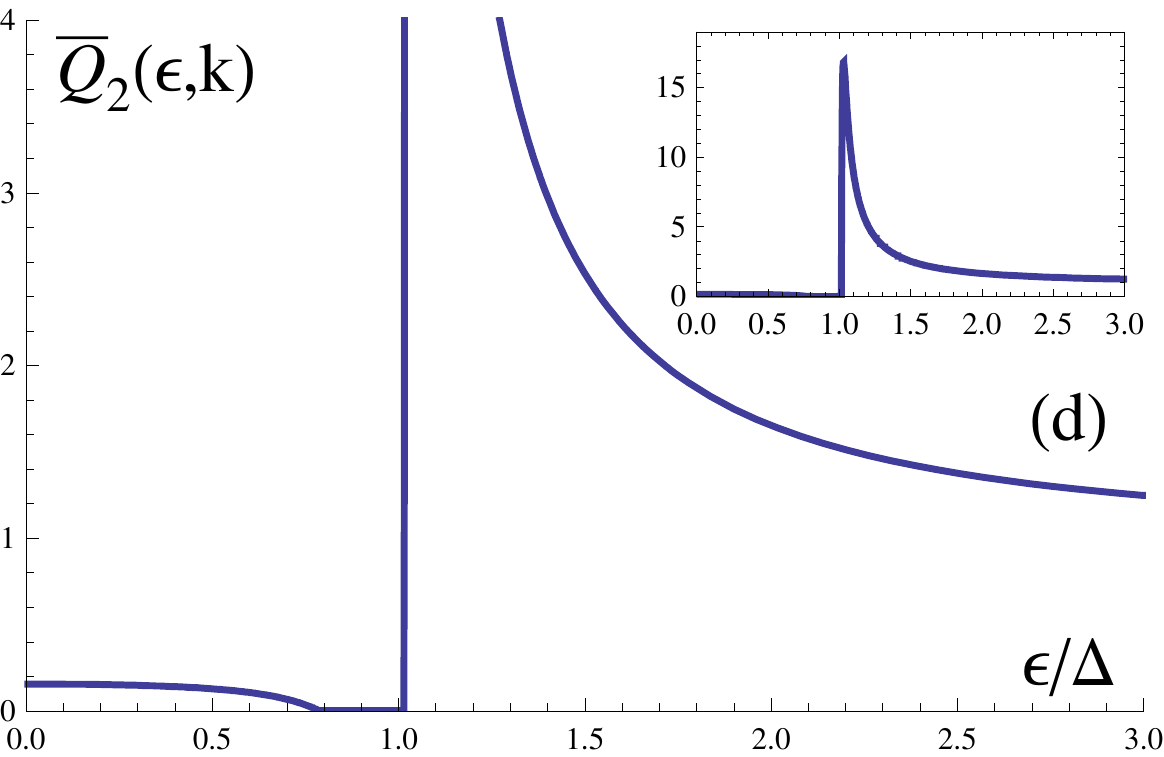}
}
\centerline{
\includegraphics[width=.70\columnwidth]{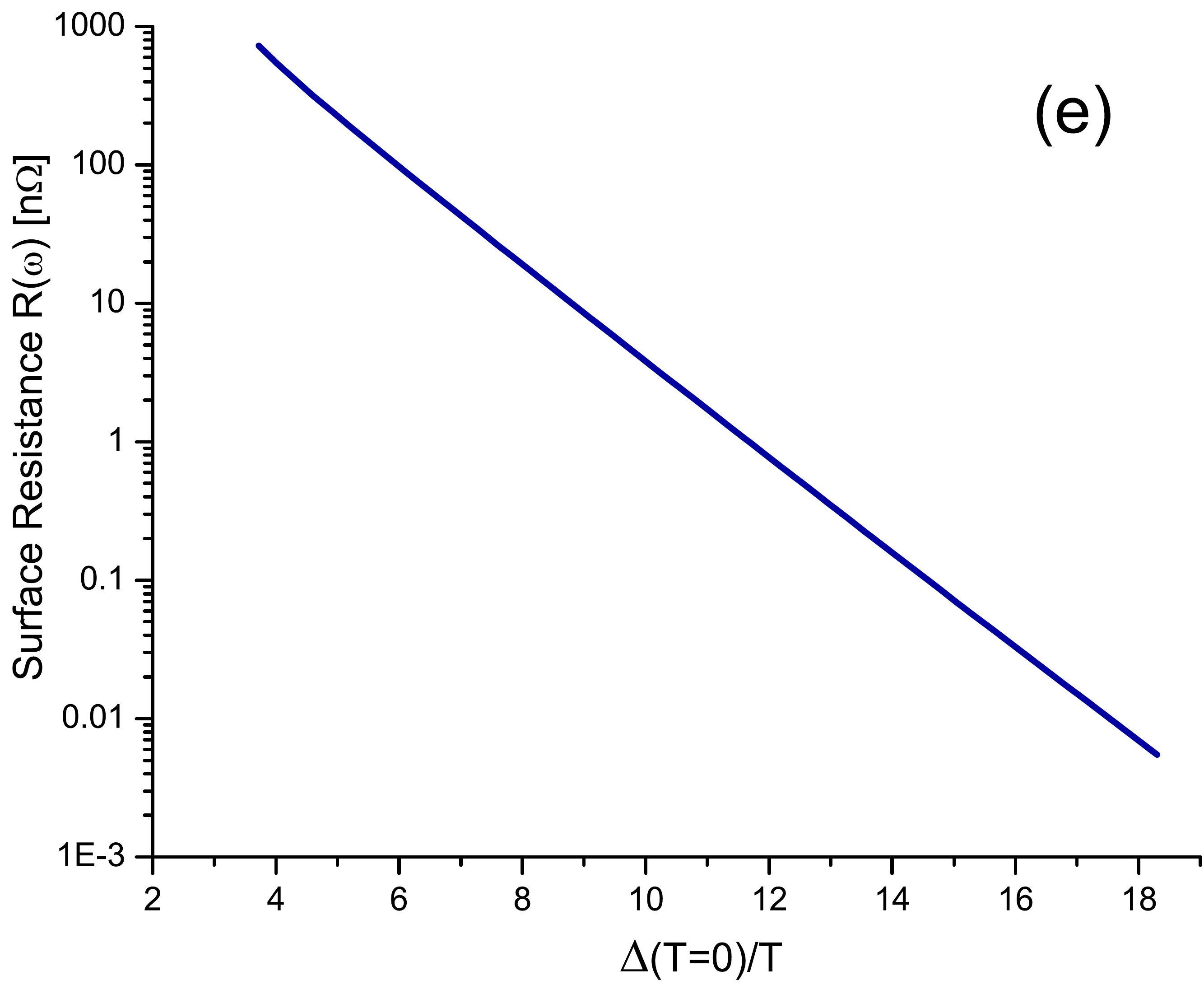}
\hspace{1cm}
\includegraphics[width=.70\columnwidth]{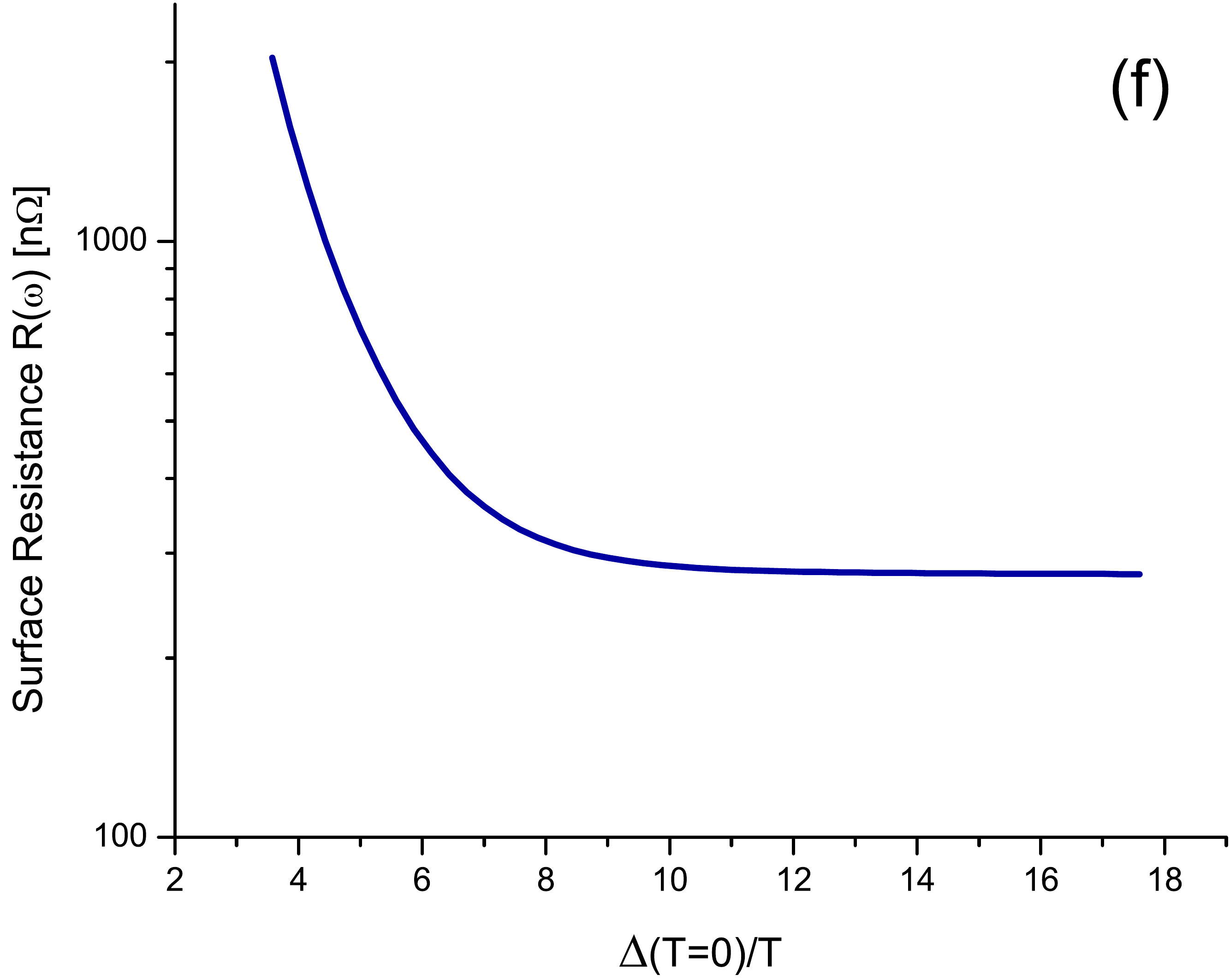}
}
\caption{
(Color online)
Regimes of ``gapped'' [(a), (c), (e), weaker exchange coupling $\nu_F J$ and lower scattering rate $1/\tau_s$;
parameters of the Shiba theory used: $1/(\tau_s \De)=0.04 $, $\ga=0.95$] and ``gapless''
[(b), (d), (f), stronger exchange coupling and/or higher scattering rate; parameters used: $1/(\tau_s\De) = 0.17$, $\ga=0$] superconductivity.
(a) and (b) The single-particle density of states (DOS) $\nu(\e)$,
obtained from the Shiba theory Eqs.~(\ref{eq:g})-(\ref{eq:ga}), and (\ref{eq:nu}).
The inset in (a) schematically shows (in red) the ``tail'' of the DOS
produced by optimal fluctuations of impurity distribution
-- exponentially small nonperturbative effect, not captured by the AG-Shiba theory and not considered in the present paper.
(c) and (d) The function $\Qbr_2(\e,k)$ [Eq.~(\ref{eq:Qb2})] describing
the dissipative contribution to the current response from a given energy $\e$,
see Secs.~\ref{sec:lowfreq} and \ref{sec:residual} for details. The plots are presented for $k=0$; the inset in (d) shows the full range of $\Qbr_2(\e,k)$.
The function $\Qbr_2(\e,k)$ is nonzero if and only if  $\nu(\e)$ is nonzero.
(e) and (f) The temperature dependence of the surface resistance $R(\om)$,
obtained from the main Eqs.~(\ref{eq:zeta0}), (\ref{eq:zeta}), (\ref{eq:Q}), (\ref{eq:jjRA}), (\ref{eq:jjRR}), and (\ref{eq:ny2D0avg})
of the paper by numerically calculating the integrals over $\e$ and $k$.
In the gapped regime (e), $R(\om) \propto \exp(-\De^*/T)$ is exponential at lower temperatures and vanishes at $T=0$.
In the gapless regime (f), for moderate ``subgap'' DOS, the surface resistance
$R(\om)$ is exponential at lower but finite temperatures and saturates to a nonzero value at $T=0$.
The latter case reproduces the commonly observed experimental behavior~\cite{review}.
}
\label{fig:main}
\end{figure*}

Magnetic impurities in
$s$-wave superconductors have been a subject of interest for a long time.
Shortly after the development of the Bardeen-Cooper-Schrieffer (BCS) theory of superconductivity~\cite{BCS},
Abrikosov and Gor'kov (AG) demonstrated~\cite{AG} that magnetic impurities introduced into the sample in moderate concentrations
lead to the suppression of superconductivity.
If the magnetic scattering rate $1/\tau_s>1/\tau_s^*$ exceeds the critical value $1/\tau_s^* \approx 0.88\, T_{c0}$,
where $T_{c0}$ is the transition
temperature of a sample without magnetic impurities (we set $\hbar=1$ throughout the paper),
the superconductivity is completely suppressed at all temperatures.
In contrast, much stronger nonmagnetic disorder is required to suppress superconductivity:
the scattering rate $1/\tau \sim \e_F$ must be on the order of the Fermi energy $\e_F \sim 10^3 T_{c0}$.

According to the AG and subsequent~\cite{LuYu,Rusinov,Shiba,Maki,Ginsberg,Balatsky,BNA,Lamacraft}
theories, even below the critical value, $1/\tau_s < 1/\tau_s^*$,
the presence of magnetic impurities can result in the regime of ``gapless superconductivity'' (GSC),
where the superconducting order parameter $\De$ is nonzero,
yet the single-particle density of states (DOS) $\nu(\e)$ does not vanish down to the Fermi
level $\e=0$,  Fig.~\ref{fig:main}.
The GSC regime is predicted to occur quite generically,
although the magnitude of the ``subgap'' DOS is parameter-dependent.
Even for low scattering rate $1/\tau_s \ll 1/\tau_s^*$
and weak exchange coupling $J$, $\nu_F J \ll 1$
($\nu_F$ is the normal state DOS at the Fermi level per one spin projection),
optimal fluctuations in the impurity distribution produce
``tails'' in the DOS~\cite{Balatsky,BNA,Lamacraft} below the ``hard gap'' predicted by the AG theory,
Fig.~\ref{fig:main}(a).
The GSC regime becomes much more pronounced with increasing the exchange coupling and/or scattering rate,
Fig.~\ref{fig:main}(b),
as the Shiba theory~\cite{Shiba,Ginsberg} demonstrates.
This is also supported by a recent numerical study~\cite{GK} of a different, but mathematically equivalent model.

In the GSC regime,  gapless quasiparticle excitations
give rise to  dissipation even at zero temperature~\cite{dissipation}.
Although this  dissipation mechanism
(caused either by the natural presence of magnetic impurities or unintentional/unavoidable contamination of the sample with them)
may be negligible for most practical applications of superconductors,
it could play an important role in devices that require high quality performance.
One example of such systems are the superconducting radio-frequency (SRF) cavities,
widely used in particle accelerators (see Ref.~\onlinecite{SRFgeneral} for a review and references therein;
another notable system is superconducting qubits).
The SRF cavities are characterized by exceptional quality factors, which are, however, limited
to a finite residual value $\sim 10^{10}$ at temperatures
$T \ll T_c$ much smaller than the superconducting transition temperature $T_c$,
where the contribution from thermally excited quasiparticles vanishes.

\begin{figure}
\includegraphics[width=.95\columnwidth]{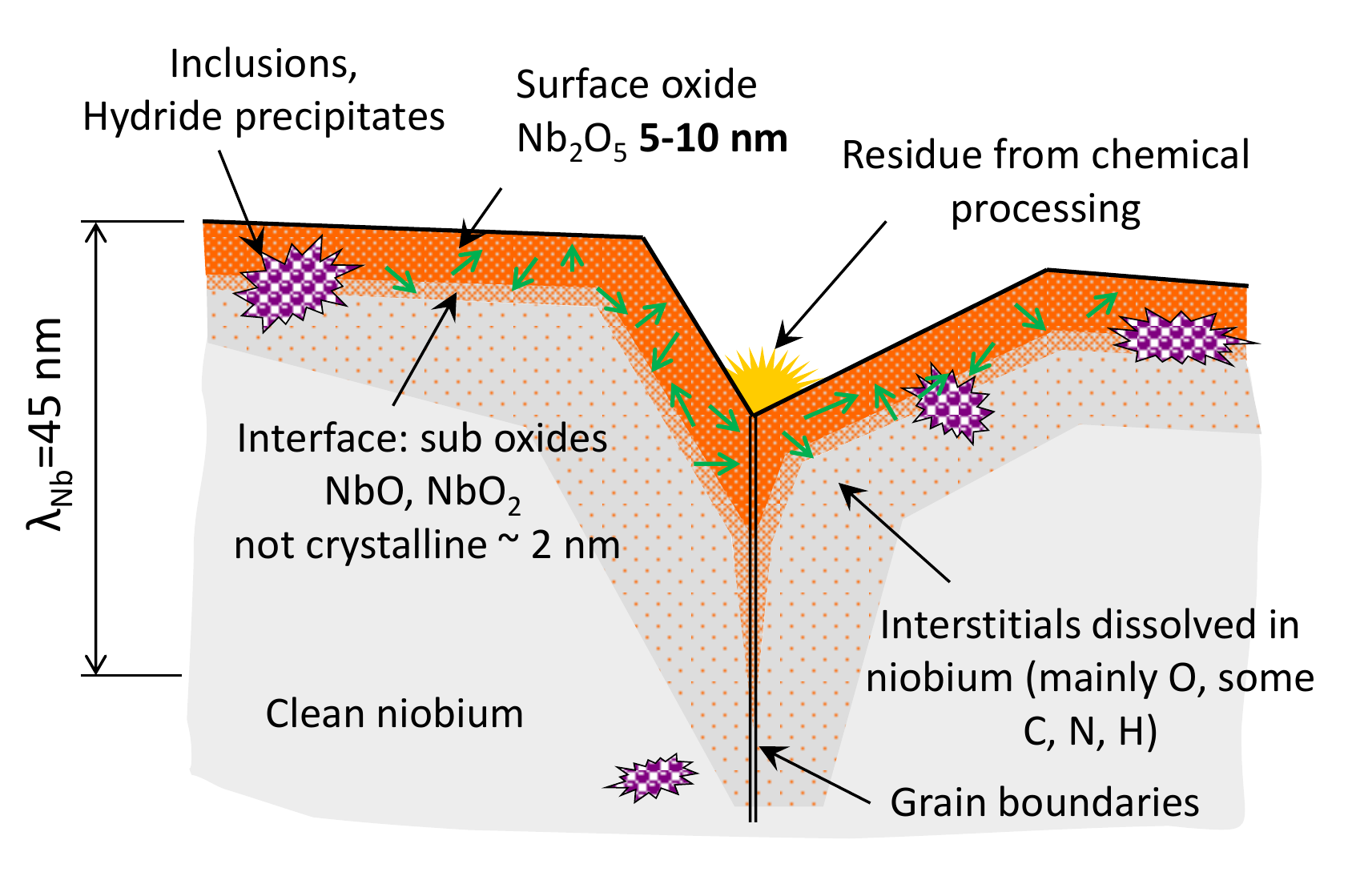}
\caption{
(Color online) Typical structure of the surface layer of the air-exposed Nb samples used for SRF cavities.
Localized magnetic moments (shown as green  arrows) can form (Ref.~\onlinecite{Cava}) in oxygen vacancies of the Nb$_2$O$_5$ layer of thickness $\sim$5-10nm.
Typical penetration depth of the radio-frequency electromagnetic field is $\sim$45nm.}
\label{fig:surfacelayer}
\end{figure}

Despite the high practical relevance of the problem,
there is no commonly accepted theoretical explanation
of the origin of the residual
Ohmic losses in SRF cavities.
Given the above properties,
it was recently argued in Ref.~\onlinecite{Proslier} that they
could  indeed be attributed to the presence of
magnetic impurities in the system.
Although the bulk of Nb samples used for SRF cavities is typically very clean,
a disordered oxide surface layer forms due to exposure to atmosphere~\cite{surflayer}, Fig.~\ref{fig:surfacelayer}.
Most importantly, magnetic moments can develop~\cite{Cava}
in the oxygen vacancies of the sub-stoichiometric Nb$_2$O$_5$ layer of thickness $\sim 5-10 \text{nm}$.
Since the typical penetration depth of the electro-magnetic field is $\sim 45 \text{nm}$,
this means that
superconductivity could be partially suppressed in a considerable fraction of the operating region.
Scanning tunneling spectroscopy measurements of the SRF cavity samples performed in Ref.~\onlinecite{Proslier}
revealed the surface tunneling DOS
with appreciable ``subgap'' contribution, considerably greater than one would expect from a high-purity Nb material.
Combined with good fits to the Shiba theory,
these data suggested
magnetic impurities in the oxide surface layer as an important contributing factor to the dissipation in SRF cavities.

To support this idea,
in the present work,
we develop a microscopic theory of the surface impedance of s-wave superconductors with magnetic impurities.
According to the surface chemistry of the air-exposed Nb samples~\cite{surflayer}, Fig.~\ref{fig:surfacelayer},
the real SRF cavity material is most appropriately described by a model of disordered surface layer
that contains both magnetic and nonmagnetic impurities,
while the rest of the sample is weakly disordered or pure.
In principle, such model can be studied in the framework of the quasiclassical approach to superconductivity
based on the Eilenberger equation~\cite{Eilenberger,Kopnin}. However,
this involves solving a self-consistency problem for a system of differential equations,
which, for realistic parameter values,
is challenging even using numerical methods.

Instead, here we consider a simpler
model of a superconducting sample
with uniform in space distribution of magnetic and nonmagnetic impurities, Fig.~\ref{fig:model}.
The main practical advantage is that for this model
we are able to analytically obtain the general
expressions for the current response function and surface impedance.
The expressions are valid, within the approximations of the theory,
in a wide range of parameter values and, in the general case,
only the resulting integrals need to be calculated numerically.
This generality allows for the application of our theory
to the description of experimental systems, such as SRF cavities and superconducting qubits,
in various practically relevant regimes.
On the other hand, if necessary, explicit analytical expressions
can be obtained in numerous limiting cases.

The current response function and surface impedance of superconductors without magnetic impurities
are provided by the Mattis-Bardeen~\cite{MB} and Abrikosov-Gor'kov-Khalatnikov~\cite{AGK}
theories (see also Ref.~\onlinecite{Halbritter,review}), in the presence and absence of nonmagnetic disorder, respectively.
The surface impedance of superconductors with magnetic impurities was previously
studied in Refs.~\onlinecite{Nam,Skalski} in the limit $\nu_F J \ll 1$ of weak exchange coupling of the AG theory.
In the present work, we consider the case of arbitrary exchange coupling strength $\nu_F J$,
treating the interactions of conduction electrons with magnetic impurities within the framework of the Shiba theory.
This sufficiently widens the range of impurity concentrations, where the GSC regime (of particular interest to us) with appreciable DOS at the Fermi level
occurs.

As the feature of key relevance to SRF cavities, we demonstrate that the presence of magnetic impurities
does lead to the saturation of the surface resistance at zero temperature in the GSC regime.

Our theory employs the linear response formalism and is therefore
valid as long as the superconducting state is not appreciably suppressed by the magnetic field $H$.
For type-II superconductors in thermal equilibrium,
the upper bound for this is set by the first critical field $H_{c1}$,
above which the system becomes unstable towards creation of vortices.
Real SRF cavities, however, are known to operate in a metastable
vortex-free state that persists up to a higher ``superheating'' field $H_\text{sh}>H_{c1}$~\cite{BeanLivingston,VL,CS,TCS}.
Thus, our theory should be applicable in the range $H \lesssim H_\text{sh}$.

At higher fields $H \sim H_\text{sh}$
one could, in fact, expect a cooperative effect of the two dissipation mechanisms:
magnetic disorder could create ``hot spot'' regions~\cite{hotspots} of locally suppressed superconductivity at the surface
and thus trigger proliferation of vortices.
Such regime deserves a separate study.

The rest of the paper is organized as follows. In Sec.~\ref{sec:model},
the studied system is presented and the main approximations are formulated.
In Sec.~\ref{sec:surfimp}, the surface impedance and current response function are introduced.
In Sec.~\ref{sec:Q}, the current response function is calculated.
In Sec.~\ref{sec:lowfreq}, the low-frequency expansion is performed.
In Sec.~\ref{sec:residual},
the key result pertaining to the presence of magnetic impurities --
finite residual surface resistance in the GSC regime -- is demonstrated.
Concluding remarks are presented in Sec.~\ref{sec:conclusion}.

\section{Model\label{sec:model}}

\begin{figure}
\includegraphics[width=.85\columnwidth]{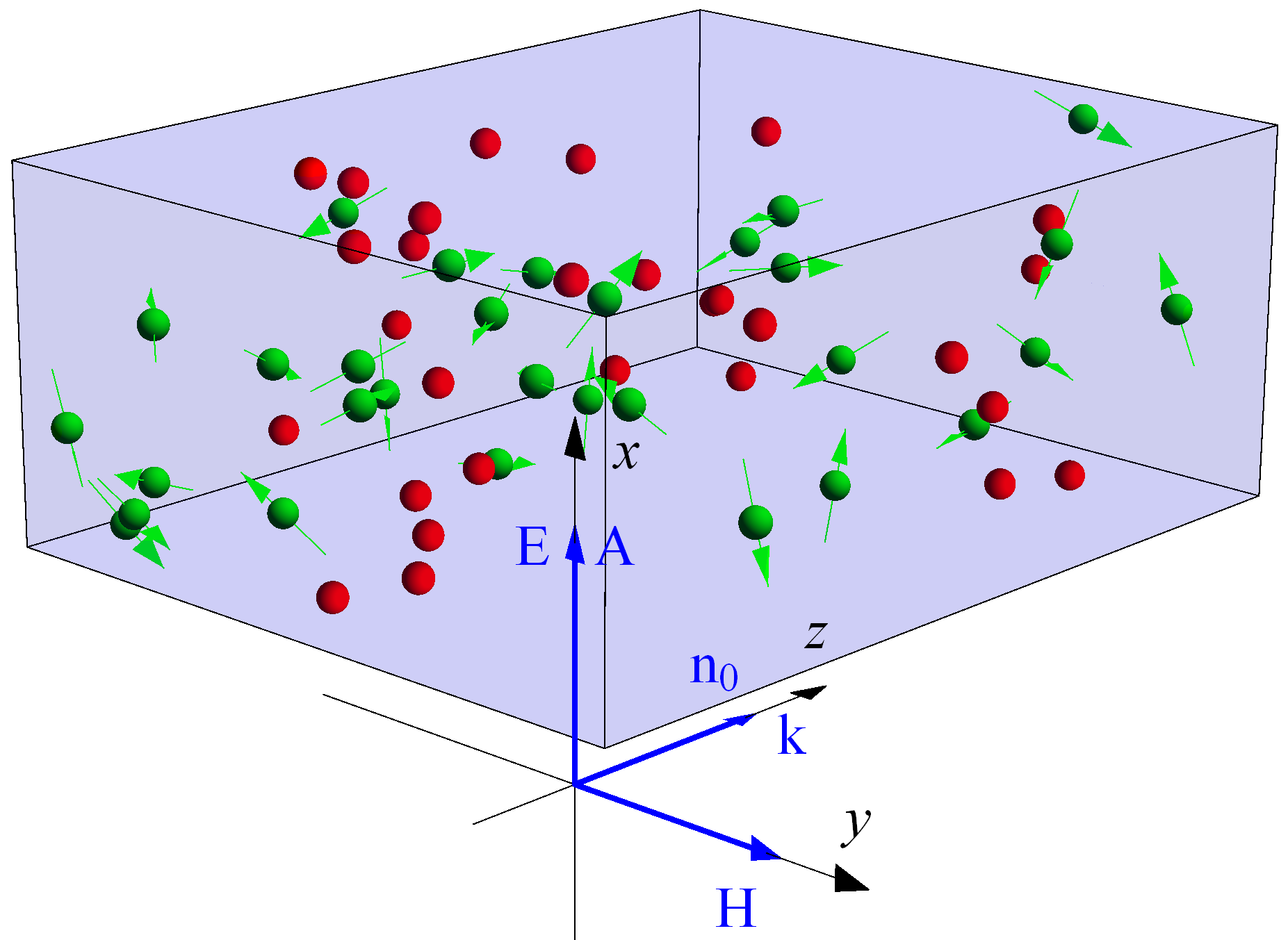}
\caption{
(Color online) Studied system: a half-infinite ($z>0$) superconducting sample
with magnetic (green) and nonmagnetic (red) impurities uniformly distributed
over its volume.
Relation between the wave-vector $\kb$, electric $\Eb$ and magnetic $\Hb$ fields, and the vector potential $\Ab$ is shown.}
\label{fig:model}
\end{figure}

We assume the superconducting sample occupies the half-space $z>0$
and contains both nonmagnetic and magnetic impurities,
which are distributed uniformly in space with densities $n$ and $n_s$, respectively, Fig.~\ref{fig:model}.
Within the framework of the BCS theory~\cite{BCS,AGD},
the Hamiltonian of the system can be written as
\newcommand{\Sb}{{\bf S}}
\bwt
\[
    \Hh= \int_{z>0} \dt^3\rb\, \lt\{ \psih^\dg_\sig \lt[E \lt(\lt|\hat{\pb} - \tfrac{e}{c} \Ab \rt|\rt)-\e_F +
        \sum\nolimits_a u \de(\rb-\rb_a)\rt] \psih_\sig + \sum\nolimits_b J \s_b (\psih^\dg_\sig \sigb^{\sig\sig'} \psih_{\sig'}) \de(\rb-\rb_b)
        +\De [\psih_\ua \psih_\da + \psih^\dg_\da \psih^\dg_\ua] \rt\}
\]
\ewt
Here, $\psih_\sig=\psih_\sig(\rb)$ is the electron field operator, $\sig, \sig'=\ua,\da$ are the spin indices,
$\sigb = (\sig_x, \sig_y, \sig_z)$ is the vector of Pauli matrices, and summation over repeated spin indices is implied;
$E(p)$ is the electron spectrum, which we assume isotropic in momentum $\pb$, $p=|\pb|$,
$\hat{\pb} = -\itxt \nabla$; $\Ab=\Ab(t,\rb)$ is the vector potential of the electromagnetic field penetrating the sample;
$\De$ is the superconducting order parameter, which
has to be found self-consistently in the presence of  magnetic impurities.

Next, $\rb_a$ and $\rb_b$ are random positions of nonmagnetic and magnetic impurities, respectively.
We use the conventional disorder averaging
technique~\cite{AGD} (``noncrossing'' approximation)
and, to keep calculations simpler, assume contact interaction potential of impurities (``point disorder'').
We (i) treat magnetic impurities as classical spins described by the unit vectors $\s_b$ and assume them unpolarized,
(ii) consider arbitrary exchange coupling strength $\nu_F J$, summing the full perturbation series for a single impurity.
These are the approximations of the Shiba theory~\cite{Shiba}.

Note that the exponentially small subgap contribution (``tail'')
to the DOS arising from the optimal fluctuations of magnetic disorder~\cite{Balatsky,BNA,Lamacraft}
is not captured within the noncrossing approximation.
This nonperturbative effect is dominant only in the limit of weak exchange coupling $\nu_F J\ll 1$
and small impurity concentration, such that $1/(\tau_s T_{c0}) \ll 1$.
In this work, we concentrate on more significant contributions
to the DOS that arise at larger impurity concentration and/or stronger exchange coupling.

\section{Surface impedance\label{sec:surfimp}}

The complex surface impedance
\beq
    Z(\om)=R(\om)+\itxt X(\om)
\label{eq:zeta0}
\eeq
(see, e.g., Refs.~\onlinecite{review,LLX})
relates the electric $\Eb(z) \etxt^{-\itxt \om t}$
and magnetic $\Hb(z) \etxt^{-\itxt \om t}$   fields at frequency $\om$
at the interface $z=0$ between the vacuum and sample as
\beq
    \Eb(0) = \frac{c}{4\pi} Z(\om) [\Hb(0) \times \nb_0],
\eeq
where $\nb_0 = (0,0,1)$ is the unit vector normal to the surface pointing into the sample, Fig.~\ref{fig:model}.
The real part $R(\om)$ of the impedance (\ref{eq:zeta0})
determines the energy flux (averaged over the oscillation period)
\[
    W= \frac{1}{2} \lt(\frac{c}{4 \pi}\rt)^2 R(\om) |\Hb(0)|^2
\]
of the electro-magnetic field per unit area
from the vacuum into the sample and is referred to as the ``surface resistance''.

The general~\cite{specrefl}  expression for the surface impedance of a uniformly disordered system
reads
\beq
    Z(\om) = -\itxt \frac{ 4 \pi \om \,\la(\om)}{c^2}, \mbox{ } \la (\om) = \frac{2}{\pi}\int_0^{+\infty} \frac{\dt k}{k^2+4 \pi Q(\om,k) /c^2}.
\label{eq:zeta}
\eeq
Here, $Q(\om,k)$ is the linear current response function of an infinite sample in Fourier representation,
dependent on the frequency $\om$ and the absolute value $k =|\kb|$ of the wave-vector $\kb$.
In Eq.~(\ref{eq:zeta}), we also introduce the complex penetration depth $\la(\om)$:
its real part $\text{Re} \la(\om)$ is the actual penetration depth that  determines the decay scale of the electromagnetic field
into the bulk. The Ohmic dissipation is determined by the imaginary part $Q_2(\om,k)$ of the
current response function
\beq
    Q(\om,k) = Q_1(\om,k) - \itxt Q_2(\om,k).
\label{eq:Q1Q2}
\eeq
According to Eq.~(\ref{eq:zeta}), the surface resistance $R(\om)$ is finite, only if $Q_2(\om,k)$ is nonzero.

The response function $Q(\om,k)$ defines the relation
\beq
    \jb (\om,\kb)  = - \frac{1}{c} Q(\om,k) \Ab (\om,\kb)
\label{eq:Qdef}
\eeq
in the Fourier representation between the electric current $\jb(\om,\kb)$ and the electro-magnetic field,
described by the vector potential $\Ab(\om,\kb)$.
The trivial tensor structure of Eq.~(\ref{eq:Qdef}) holds for cubic crystal symmetry
and, in particular, for the isotropic electron spectrum $E(|\pb|)$ assumed here.
It is convenient to work in the gauge of absent scalar potential $\vphi(\om,\kb)=0$.
Additionally, the vector potential $\Ab (\om,\kb)$ may be assumed to satisfy the constraint
\beq
    \Ab(\om,\kb) \kb = 0,
\label{eq:Ak0}
\eeq
which significantly simplifies the calculations.
This is equivalent to the local electroneutrality condition, which is a very good approximation for superconductors.
The geometric relation between the wave-vector $\kb$, vector potential $\Ab(\om,\kb)$,
and electric $\Eb(\om,\kb) = \frac{\itxt \om}{c} \Ab(\om,\kb)$
and magnetic $\Hb(\om,\kb) = [\itxt \kb \times \Ab(\om,\kb)]$ fields is shown in Fig.~\ref{fig:model}.

In the next section, we calculate the current response function $Q(\om,k)$,
which fully determines the surface impedance (\ref{eq:zeta}).

\section{Current response function\label{sec:Q}}

According to the general Kubo formalism~\cite{AGD,LLX}, the expression for the linear current response function
[Eq.~(\ref{eq:Qdef})] can be written down as
\bwt
\beq
    Q(\om,k) = Q_0\frac{-3 \itxt}{2} \int_{-\infty}^{+\infty} \dt \e
    \lt[\lt( \tanh \frac{\e_+}{2 T} - \tanh \frac{\e_-}{2 T} \rt) \lan j j \ran^{RA}(\e,\om,k)
     +\tanh \frac{\e_-}{2 T}  \lan j j \ran^{RR}(\e,\om,k)
    - \tanh \frac{\e_+}{2 T}  \lan j j \ran^{AA}(\e,\om,k)\rt],
\label{eq:Q}
\eeq
\beq
   \lan j j \ran^{ab} (\e,\om,k) = \lt\lan n_\al^2 \int \frac{\dt \xi}{2 \pi}
        \lt[G^a(\e_+,\pb_+) G^b(\e_-,\pb_-)
        + F^a (\e_+,\pb_+) F^b(\e_-,\pb_-)\rt] \rt\ran_\nb, \mbox{ } a,b=R,A.
\label{eq:jjdef}
\eeq
\ewt

In Eq.~(\ref{eq:Q}), we introduced
\beq
    Q_0 = 2 (e v_F)^2 \nu_F/3,
\label{eq:Q0}
\eeq
where $v_F = (\dt E/\dt p)_{p=p_F}$ is the Fermi velocity,
$p_F$ is the Fermi momentum, $E(p_F) = \e_F$, and $\nu_F = p_F^2 /(2 \pi^2 v_F)$.
The quantity $Q_0=Q(0,0)|_{T=0,\text{clean}}$ is the value of the current response function for a clean system at $\om=0$, $k=0$, and $T=0$;
it can be related to the formal London penetration depth $\la_{L0}$
(which can be introduced as a characterization parameter, regardless whether the London limit actually applies)
as
\beq
    1/\la_{L0}^2= 4 \pi Q_0/c^2.
\label{eq:laL0}
\eeq
The quantities $Q_0$ and $\la_{L0}$ describe
the corresponding clean system and are determined only by the band structure parameters.

In Eq.~(\ref{eq:jjdef}),
\beq
    G^R(\e,\pb) = \frac{\et+\xi}{\et^2-\xi^2-\Det^2},
    F^R(\e,\pb) = \frac{\Det}{\et^2-\xi^2-\Det^2},
\label{eq:GF}
\eeq
\[
G^A(\e,\pb) = [G^R(\e,\pb)]^*, \mbox{ }
F^A(\e,\pb) = [F^R(\e,\pb)]^*,
\]
are the retarded ($R$)
and advanced ($A$) ``normal'' ($G$) and ``anomalous'' ($F$) Green's functions, averaged over disorder,
where the conventionally introduced~\cite{AG,Shiba}
functions $\et$ and $\Det$ are defined below.
For point disorder and due to the property (\ref{eq:Ak0}),
the ``ladder'' contribution~\cite{AGD}
vanishes, and the current-current correlation functions (\ref{eq:jjdef}) are determined
by the products of disorder-averaged Green's functions.

Further, in Eqs.~(\ref{eq:Q}), (\ref{eq:jjdef}), and (\ref{eq:GF}),
$\e$ is the energy relative to the Fermi level $\e_F$, and
$\e_\pm = \e \pm \om/2$, $\pb_\pm = \pb \pm \kb/2$.
We split the integration
over momentum
\[
    \int \frac{\dt^3 \pb}{(2 \pi)^3} \ldots = \nu_F \lt\lan \int \dt \xi ... \rt\ran_\nb,
    \mbox{ }
    \lan \ldots \ran_\nb =\int_{|\nb|=1} \frac{\dt \nb}{4 \pi} \ldots,
\]
in a standard way into the integration over its absolute value $p=|\pb|$,
expressed in terms of $\xi = E(p) - \e_F$, and averaging over its direction, expressed in terms of the unit vector $\nb = \pb/p$.

Equation~(\ref{eq:jjdef}) defines the correlation functions
of the current components perpendicular to $\kb$, see Eq.~(\ref{eq:Ak0}),
and so, $n_\al$ are the components of $\nb=(n_x,n_y,n_z)$ perpendicular to $\kb$:
if $\kb= (0,0,k)$, as in Fig.~\ref{fig:model}, then $\al=x,y$.
Since the spectrum is isotropic,
the integral $I(\nb \kb) = \int \frac{\dt \xi}{2\pi} \ldots$  in Eq.~(\ref{eq:jjdef})
depends just on $\nb \kb = n_z k$ and
angular averaging is reduced to calculating the integral
\beq
    \lan n_\al^2 I(\nb \kb) \ran_\nb = \frac{1}{4} \int_{-1}^1 \dt n_z (1-n_z^2) I(n_z k).
\label{eq:angavg}
\eeq

Note that for
the integration order as in Eqs.~(\ref{eq:Q}) and (\ref{eq:jjdef}) -- first over $\xi$ and then over $\e$ --
the contribution to $Q(\om,k)$ arising from the dependence of the current operator on the vector potential is already compensated for.

It is convenient to introduce the (retarded) quasiclassical Green's functions~\cite{Eilenberger,Kopnin}
\beq
    \{g,f\}(\e) = \frac{\itxt}{\pi} \int \dt \xi \{G^R,F^R\} (\e,\pb),
\label{eq:gfdef}
\eeq
which, according to Eq.~(\ref{eq:GF}), equal
\beq
    g(\e) = \frac{\et}{\sqrt{\et^2-\Det^2}}= \frac{v}{\sqrt{v^2-1}} ,
\label{eq:g}
\eeq
\beq
    f(\e) = \frac{\Det}{\sqrt{\et^2-\Det^2}}= \frac{1}{\sqrt{v^2-1}}.
\label{eq:f}
\eeq

Within the Shiba theory~\cite{Shiba}, the function
\beq
    v=v(\e) \equiv   \frac{\tilde{\epsilon}}{\Det} = \frac{g(\e)}{f(\e)}
\label{eq:vdef}
\eeq
satisfies the equation
\beq
    v \De = \e  + \frac{1}{\tau_s} \frac{\sqrt{1-v^2}}{\ga^2-v^2} v.
\label{eq:v}
\eeq
Here,
\beq
    \ga =  \frac{1-(\pi \nu_F J)^2}{1+( \pi \nu_F J)^2}
\label{eq:ga}
\eeq
is the parameter of the Shiba theory characterizing exchange coupling strength and $\tau_s$ is the scattering time on magnetic impurities,
\[
    \frac{1}{\tau_s} = \frac{n_s}{2 \pi \nu_F} (1-\ga^2)  =  2 \pi \nu_F n_s \frac{J^2}{[1+(\pi \nu_F J)^2]^2}.
\]
In the weak coupling limit $\nu_F J \ll 1$ of the AG theory~\cite{AG},  $\ga^2=1$.

We also introduce the function $h(\e)$ as
\beq
    \sqrt{\et^2-\Det^2} = h(\e) +\frac{\itxt}{2\tau},
\label{eq:hdef}
\eeq
which is related to $g(\e)$ and $f(\e)$ as
\beq
    h(\e) = \frac{1}{2} \lt( \frac{\e}{g(\e)} + \frac{\De}{f(\e)} \rt).
\label{eq:h}
\eeq
In Eq.~(\ref{eq:hdef}), $\tau$ is the scattering time on nonmagnetic impurities,
\[
    \frac{1}{\tau} = 2 \pi \nu_F n u^2.
\]

Solving Eq.~(\ref{eq:v}) for $v$ (in the general case -- numerically), one obtains $g(\e)$, $f(\e)$, and $h(\e)$.
Integration over $\xi$ in Eq.~(\ref{eq:jjdef}) is straightforward and we obtain
\bwt
\beq
    \lan j j \ran^{RA}(\e,\om,k)  = \frac{1}{2}[g(\e_+) g^*(\e_-) + f(\e_+) f^*(\e_-) +1 ]
        \lan \overline{jj}\ran^{RA}(\e,\om,k),
        \mbox{ }
            \lan \overline{jj} \ran^{RA}(\e,\om,k) = \lt\lan \frac{\itxt n_\al^2}{h(\e_+)-h^*(\e_-) +\itxt/\tau - v_F \nb \kb } \rt\ran_\nb,
\label{eq:jjRA}
\eeq
\beq
     \lan j j\ran^{RR}(\e,\om,k) =
    \frac{1}{2} [1-g(\e_+) g(\e_-) - f(\e_+) f(\e_-)]
        \lan \overline{jj}\ran^{RR}(\e,\om,k),
        \mbox{ }
            \lan \overline{jj} \ran^{RR}(\e,\om,k) = \lt\lan \frac{\itxt n_\al^2}{h(\e_+) + h(\e_-) + \itxt/\tau - v_F \nb \kb } \rt\ran_\nb,
\label{eq:jjRR}
\eeq
\ewt
and $ \lan jj\ran^{AA}(\e,\om,k) = [\lan jj \ran^{RR}(\e,\om,k)]^*$,
$ \lan \overline{jj}\ran^{AA}(\e,\om,k) = [\lan \overline{jj} \ran^{RR}(\e,\om,k)]^*$.
Angular averaging in Eq.~(\ref{eq:jjRA}) and (\ref{eq:jjRR}) can also be performed explicitly according to Eq.~(\ref{eq:angavg}),
\beq
        \lan \overline{jj}\ran^{RR,RA}(\e,\om,k)
        =\frac{-\itxt}{4 v_F k}  \lt[\lt(1-l_0^2\rt) \ln \frac{l_0 - 1}{l_0+ 1} - 2 l_0\rt],
\label{eq:ny2D0avg}
\eeq
where
\[
    l_0=  \frac{1}{v_F k} \times \lt\{ \begin{array}{ll}
            h(\e_+) - h^*(\e_-) + \itxt/\tau, & \mbox{for } RA, \\
            h(\e_+) + h(\e_-) + \itxt/\tau, & \mbox{for } RR.
            \end{array}\rt.
\]

Equations (\ref{eq:Q}), (\ref{eq:jjRA}), (\ref{eq:jjRR}), and (\ref{eq:ny2D0avg}),
combined with Eqs.~(\ref{eq:g})-(\ref{eq:h}) of the Shiba theory,
provide the answer for the current response function $Q(\om,k)$
and constitute the main result of our work.
Within the approximations of the theory, these equations are valid
at arbitrary values of frequency $\om$, temperature $T$, and six microscopic parameters characterizing the system.
Three standard parameters~\cite{MB,AGK} describe the clean system:
(i) the superconducting transition temperature $T_{c0}$ or, equivalently, the superconducting order parameter $\De_0$ at $T=0$
for a system without magnetic impurities;
(ii) the current response $Q_0$ [Eq.~(\ref{eq:Q0})] or, equivalently, the formally introduced  London penetration depth $\la_{L0}$ [Eq.~(\ref{eq:laL0})]
for a clean system at $T=0$;
and (iii) the Fermi velocity $v_F$.
The other three parameters describe disorder:
(i) the nonmagnetic scattering time $\tau$;
(ii) the magnetic scattering time $\tau_s$;
and (iii) the exchange coupling strength $\nu_F J$ or, equivalently, the Shiba parameter $\ga$ [Eq.~(\ref{eq:ga})].

In the general case,
Eqs.~(\ref{eq:Q}), (\ref{eq:jjRA}), (\ref{eq:jjRR}), and (\ref{eq:ny2D0avg})
provide the most explicit analytical form of the current response function
$Q(\om,k)$ possible. The function is given by the integral over energy $\e$ in Eq.~(\ref{eq:Q}), where
the dependence of the integrand on the absolute value of momentum $k$ is explicit in Eqs.~(\ref{eq:ny2D0avg}),
while the dependence on $\e$ is obtained from the well-known Shiba equation (\ref{eq:v}),
the solution to which determines the functions $g(\e)$, $f(\e)$, and $h(\e)$.

For arbitrary values of parameters, the solution to the Shiba equation
and the integrations over $\e$ for the current response $Q(\om,k)$
and over $k$ for the surface impedance $Z(\om)$ [Eq.~(\ref{eq:zeta})]
need to be carried out numerically.
The generality of the obtained results, however, should make them applicable to a variety of realistic experimental regimes.

On the other hand, in a number of limiting cases, the general
formulas can be further simplified and in many cases explicit analytical expressions
for the current response function and surface impedance can be obtained.
Analysis of such limiting cases is straightforward and we do not present it here.

In the weak coupling limit $\nu_F J \ll 1$ of the AG theory~\cite{AG} ($\ga=1$),
the results of Refs.~\onlinecite{Nam,Skalski} are (presumably) recovered,
and in the complete absence of magnetic impurities [$1/\tau_s=0$,
$h(\e) = \sqrt{\e^2-\De^2}$,
$g(\e) =\e/h(\e)$,
$f(\e) =\De/h(\e)$]
the Mattis-Bardeen theory~\cite{MB} is reproduced.

In the next two sections, we consider the low-frequency limit $\om \ll \De$, most relevant for practical applications to SRF cavities,
and concentrate on the key property pertaining to the presence of magnetic impurities
-- finite residual surface resistance in the GSC regime.

\section{Low frequency expansion \label{sec:lowfreq}}

The typical operating frequencies of the SRF cavities are $\om \sim c/L \sim 10^{-2}\text{meV} \ll \De \sim 1 \text{meV}$,
where $L \sim 10\text{cm}$ is the typical size of the cavity.
Thus for practical applications to SRF cavities, one may perform the low-frequency expansion in $\om \ll \De$.
Separating the real (``nondissipative'') and imaginary (``dissipative'') parts~[Eq.~(\ref{eq:Q1Q2})]
of $Q(\om,k)$ [Eq.~(\ref{eq:Q})],
in the leading order in $\om$ for each, we obtain,
\bwt
\beq
    Q_1(k) = Q_0 \frac{3\itxt}{2} \int_{-\infty}^{+\infty} \dt \e
    \tanh \frac{\e}{2 T}  \lt\{  [f(\e)]^2 \lan\overline{jj}\ran^{RR}(\e,0,k)
            - [f^*(\e)]^2 \lan\overline{jj}\ran^{AA}(\e,0,k) \rt\},
\label{eq:Q1}
\eeq
\beq
    Q_2(\om,k) = Q_0 \om \int_{-\infty}^{+\infty} \dt \e\, \lt(  - \frac{\dt n_0(\e)}{\dt \e} \rt)
\bar{Q}_2(\e,k),
\label{eq:Q2}
\eeq
\beq
    \bar{Q}_2(\e,k) = \frac{3}{2}\lt\{
    [f(\e)]^2
\lan\overline{jj}\ran^{RR}(\e,0,k)
       +
        [f^*(\e)]^2
\lan\overline{jj}\ran^{AA}(\e,0,k)
+
      [1+ |g(\e)|^2+ |f(\e)|^2]
\lan\overline{jj}\ran^{RA}(\e,0,k)
\rt\}.
\label{eq:Qb2}
\eeq
\ewt
Here, $n_0(\e) = 1/[\exp(\e/T)+1]$ is the Fermi distribution function.

The real part $Q_1(k)$ [Eq.~(\ref{eq:Q1})]
is finite at $\om=0$ and determines the penetration depth $\la(\om=0)$ [Eq.~(\ref{eq:zeta})] of the quasistatic magnetic field (Meissner effect).
On the other hand,
the imaginary part $Q_2(\om,k) \propto \om $ [Eqs.~(\ref{eq:Q2}) and (\ref{eq:Qb2})], which
determines the dissipation,
is nonzero only at finite frequency and is linear in it~\cite{linom} at $1/\tau_s \gg \om$.
Since $Q_2(\om,k)$ is smaller than $Q_1(k)$ at least in $\om/\De$, one may also expand Eq.~(\ref{eq:zeta})
in $Q_2(\om,k)$ to obtain
\beq
    Z(\om) = \frac{32 \,\pi \om}{c^4} \int_0^{+\infty} \dt k \frac{Q_2(\om,k)}{\lt[ k^2 + 4\pi Q_1(k)/c^2 \rt]^2}.
\label{eq:zetalowom}
\eeq
Thus the surface resistance $R(\om) \propto \om^2$ is quadratic in frequency,
which is the most common dependence observed experimentally in SRF cavities~\cite{SRFgeneral}.

\section{Residual surface resistance \label{sec:residual}}

We now turn to the key finding of our work.
The function $\bar{Q}_2(\e,k)$ [Eqs.~(\ref{eq:Qb2})] describes the contribution to the dissipative part $Q_2(\om,k)$ [Eq.~(\ref{eq:Q2})]
of the current response function
from quasiparticles with a given energy $\e$,
while the derivative $-\dt n_0(\e) /\dt \e$
constrains their distribution to the range $|\e| \lesssim T$ around the Fermi level.
As is well known~\cite{AGD,Kopnin}, the real part $g_1(\e)$ of the
normal Green's function $g(\e) = g_1(\e) - \itxt g_2(\e)$ [Eq.~(\ref{eq:g})] determines the DOS
\beq
    \nu(\e)= \nu_F g_1(\e).
\label{eq:nu}
\eeq
Inspecting Eqs.~(\ref{eq:Q2}) and (\ref{eq:Qb2}),
we notice that
\beq
    \bar{Q}_2(\e,k)=0 \Leftrightarrow  \nu(\e)=0.
\label{eq:Qb20}
\eeq
Indeed, if $g_1(\e)=0$, i.e., $g(\e) = - \itxt g_2(\e)$ is imaginary,
then, according to Eqs.~(\ref{eq:g}), (\ref{eq:f}), and (\ref{eq:h}),
so are $f(\e)= -\itxt f_2(\e)= -\itxt \sqrt{1+g_2^2(\e)}$ and $h(\e)$.
In this case,
$\lan\overline{jj}\ran^{RA}(\e,0,k)=
\lan\overline{jj}\ran^{RR}(\e,0,k)=
\lan\overline{jj}\ran^{AA}(\e,0,k)$
and the function (\ref{eq:Qb2}) does vanish,
\[
   \bar{Q}_2(\e,k) = \frac{3}{2}
   \lan\overline{jj}\ran^{RA}(\e,0,k)
[ -2 f_2^2(\e)+1+g_2^2(\e)+f_2^2(\e)]=0.
\]
We do not present a more cumbersome rigorous proof of the converse here.
Instead, the property (\ref{eq:Qb20}) is clearly illustrated in Figs.~\ref{fig:main}(a), (b), (c), (d).

Thus, as one would intuitively expect, only the energies $\e$ at which the DOS $\nu(\e)$ is nonzero contribute to dissipation.
At $T=0$ the envelope function $-\dt n_0(\e)/\dt \e \rtarr \de(\e)$ becomes a delta-function
and only the excitations at the Fermi level $\e=0$ contribute,
\beq
    Q_2(\om,k)|_{T=0} =  Q_0 \om \bar{Q}_2(\e=0,k).
\label{eq:Q1T0}
\eeq
According to~(\ref{eq:Qb20}), as the central result,
we obtain that the system exhibits finite surface resistance (\ref{eq:zetalowom}) at $T=0$
if and only if the DOS at the Fermi level is nonvanishing, i.e. the system is in the GSC regime.
\[
    R(\om)|_{T=0}>0 \Leftrightarrow \nu(\e=0)>0.
\]

The result is illustrated in Figs.~\ref{fig:main}.
If the spectrum has a gap $\bar{\De}$ (DOS ``tails''~\cite{Balatsky,BNA,Lamacraft}
are not captured by the AG-Shiba theory),
Figs.~\ref{fig:main}(a),(c),(e),
the surface resistance obeys an activation law $R(\om) \propto \etxt^{-\bar{\De}/T}$ at temperatures $T\ll \bar{\De}$, eventually
vanishing at $T=0$.

On the other hand, in the gapless regime, Figs.~\ref{fig:main}(b),(d),(f),
the surface resistance $R(\om)$ saturates to a finite value at $T=0$.
We note that in experiments it is quite typical~\cite{review} for $R(\om)$
to exhibit both the saturation at $T=0$
and an activation behavior $R(\om) \propto \exp(-\De^*/T)$
at finite but low temperatures $T \lesssim \De^*$,
with $\De^*$ close to the value of the superconducting order parameter $\De$
of a clean sample.
Such behavior can be realized, if the finite DOS $\nu(\e=0) \ll \nu_F$ at the Fermi level is much smaller
than the DOS $\nu(\e \gtrsim \De) \sim \nu_F$ above the ``nominal'' gap.
In the framework of the Shiba theory,
this is possible in the limit of low magnetic scattering rate $1/\tau_s \ll \De$ and stronger exchange coupling $\nu_F J \sim 1$,
the case shown in Figs.~\ref{fig:main}(b),(d),(f).
Thus, our microscopic model can reproduce the typical experimental temperature dependence of the surface resistance.

\section{Conclusion\label{sec:conclusion}}

In conclusion, we developed a microscopic analytical theory of the surface impedance of $s$-wave superconductors
with magnetic impurities.
The theory can potentially be applied to a variety of superconducting systems
and is of direct relevance to the problem of the residual surface resistance of SRF cavities.
We explicitly demonstrated that, in the regime of gapless superconductivity,
the system exhibits saturation of the surface resistance at zero temperature -- a routinely observed,
but largely unexplained experimental feature.
This substantiates  the recent conjecture~\cite{Proslier}
that magnetic impurities, formed at the surface of the oxide surface layer,
could be the dominant dissipation mechanism limiting the performance of the SRF cavities.
Our theory is valid in the wide range of parameter values and can be used for direct comparison with experimental data,
as will be presented elsewhere~\cite{prep}.

{\em Note added.} After a preprint~\cite{arxiv} of the present work became available,
a paper~\cite{Fominov} came out, in which the same problem
was studied in the diffusive limit ($\tau \De \ll 1$) and for weak exchange coupling ($\nu_F J \ll 1$),
with an emphasis on the quasiparticle contribution from the vicinity of the spectral gap.
The approach of Ref.~\onlinecite{Fominov}
based on the Keldysh formalism and Usadel equation is equivalent to ours in that limit and
most of the analytical results of Ref.~\onlinecite{Fominov} can be obtained
as asymptotic expansions of our general formulas.

\section{Acknowledgements}

We are thankful to A. Gurevich, A. Koshelev, G. Ciovati, and J. Zasadzinski for insightful discussions.
This work was supported by the U.S. DOE under Contract No. DE-AC02-06CH11357 at ANL;
M.K. was also supported by the U.S. DOE under Contract No. DE-FG02-99ER45790 at Rutgers.


\begin{thebibliography}{99}
\bibitem{dissipation}
Even if the subgap states are localized, dissipation is finite at nonzero frequency -- the case we are interested in --
according to the Mott's mechanism.
\bibitem{specrefl}
Although Eq.~(\ref{eq:zeta}) was derived~\cite{RS,Pippard}
for specular reflection of electrons at the interface with vacuum,
this assumption is generally not crucial:
since the current is created predominantly by electrons with momenta nearly parallel to the surface,
the impedance is only weakly sensitive to the properties of the surface scattering~\cite{RS,Pippard}.
\bibitem{linom}
Note that in the absence of magnetic impurities, $1/\tau_s=0$, the integral over $\e$ in Eq.~(\ref{eq:Q2})
has a logarithmic divergency at $\e=\De$, which has to be cut by frequency $\om$.
However, at finite $1/\tau_s$, the integral is convergent.
\end{thebibliography}
\end{document}